\documentclass[letterpaper, 10 pt, conference]{ieeeconf}  %

\usepackage{nicematrix}
\usepackage{tikz}
\usetikzlibrary{calc,shadings,patterns,tikzmark}
\usepackage{nicematrix}
\usepackage{changepage}
\usepackage{capt-of}
\usepackage[nospace]{cite}
\usepackage{balance}

\usepackage{hyperref}
\hypersetup{
    colorlinks=true,    
    urlcolor=cyan,
    pdfpagemode=FullScreen,
   }

\usepackage{graphicx}
\usepackage{multicol}                  %
\usepackage{pgfplotstable}             %
\usepackage{pgfplots}                  %
\pgfplotsset{compat=1.8}               %
\usepackage{xspace}
\usepackage{comment}
\usepackage{csquotes}                  %
\usepackage{xcolor}
\usepackage{cuted}

\definecolor{myGrey}{RGB}{78, 80, 92}
\definecolor{myLightGrey}{RGB}{230, 230, 230}
\definecolor{myMidLightGrey}{RGB}{150, 150, 150}
\definecolor{myDarkGreen}{RGB}{42, 74, 33}

\definecolor{myBlue}{HTML}{66A3F2}
\definecolor{myGreen}{HTML}{6FBFA2}
\definecolor{myRed}{HTML}{F27777}
\definecolor{myYellow}{HTML}{F2CF8D}

\makeatletter
\DeclareRobustCommand\onedot{\futurelet\@let@token\@onedot}
    \def\@onedot{\ifx\@let@token.\else.\null\fi\xspace}
    \def\eg{\emph{e.g}\onedot}

    \def\etal{\emph{et al}\onedot}
    
    \makeatother
    
\newcommand{\CCJ}{\textit{Classical Jogging}\xspace}    
\newcommand{\CDC}{\textit{Direct Control}\xspace}           
\newcommand{\CDCwG}{\textit{Gripper Control}\xspace}   
\newcommand{\NCCJ}{Classical Jogging\xspace}     
\newcommand{\NCDC}{Direct Control\xspace}         
\newcommand{\NCDCwG}{Gripper Control\xspace}

\IEEEoverridecommandlockouts                              %

\title{\LARGE \bf
Optimizing Robot Programming: Mixed Reality Gripper Control 
}

\author{Maximilian Rettinger$^{*}$, Leander Hacker, Philipp Wolters, Gerhard Rigoll \\ 
Technical University of Munich %
\thanks{$^{*}$ e-mail: maximilian.rettinger@tum.de}%
}

\begin{document}

\maketitle
\thispagestyle{empty}
\pagestyle{empty}

\begin{strip}
\begin{minipage}{\textwidth}\centering
\vspace{-30pt}
\includegraphics[width=1.0\textwidth]{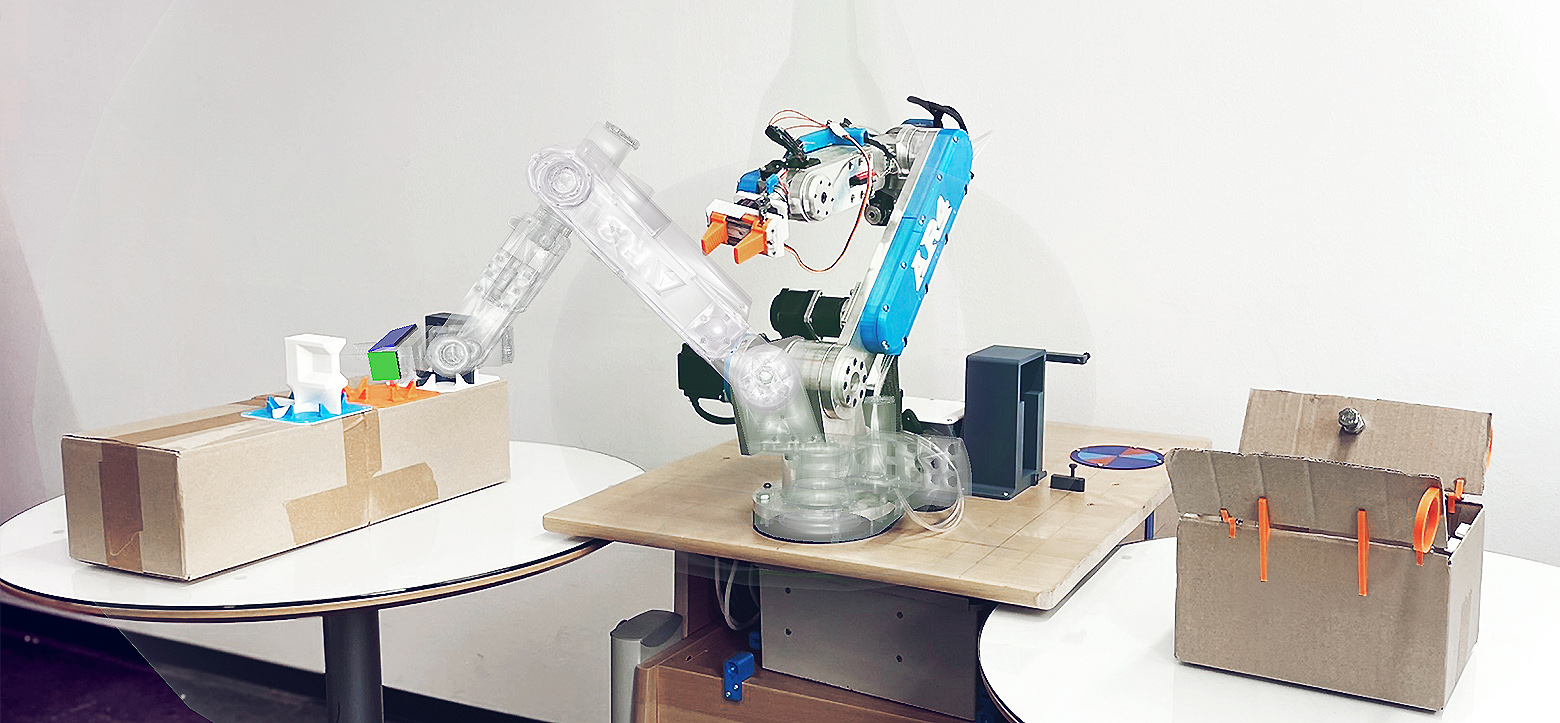} %
\captionof{figure}{Illustration of the physical and virtual robot arm. The transparent shape represents the working envelope. The boxes on the left and right are used for the user study tasks.}%
\label{figureTeaser}
\end{minipage}
\end{strip}

\begin{abstract}
    Conventional robot programming methods are complex and time-consuming for users. In recent years, alternative approaches such as mixed reality have been explored to address these challenges and optimize robot programming. While the findings of the mixed reality robot programming methods are convincing, most existing methods rely on gesture interaction for robot programming. Since controller-based interactions have proven to be more reliable, this paper examines three controller-based programming methods within a mixed reality scenario: 1) \CCJ, where the user positions the robot's end effector using the controller's thumbsticks, 2) \CDC, where the controller's position and orientation directly corresponds to the end effector's, and 3) \CDCwG, where the controller is enhanced with a 3D-printed gripper attachment to grasp and release objects. A within-subjects study ($n=30$) was conducted to compare these methods. The findings indicate that the \CDCwG condition outperforms the others in terms of task completion time, user experience, mental demand, and task performance, while also being the preferred method. Therefore, it demonstrates promising potential as an effective and efficient approach for future robot programming. 
    Video available at \href{https://youtu.be/83kWr8zUFIQ}{https://youtu.be/83kWr8zUFIQ}.

\end{abstract}

\section{INTRODUCTION}
    Mixed reality (MR) \cite{milgramMR} allows users to augment the physical environment with virtual elements, offering promising opportunities in fields ranging from healthcare \cite{medicalMR, MedicalMR2, NatureSrDialysis} to ordnance disposal \cite{eodMR, EodMR2}. 
    For robot programming, this potential has been explored for several years \cite{historyARrobotProgramming3, historyARrobotProgramming2, historyARrobotProgramming1} since traditional programming approaches are unintuitive, highly skill-demanded, or time-consuming \cite{traditionalProgrammingNegative1, traditionalProgrammingNegative2}. 
    For instance,  in online programming, the robot is moved to a position via a teach pendant by jogging per cycle, which is tedious and time-consuming \cite{conventionalOnlineTeachPendants}. In offline programming, the task is created on an independent computer, which provides features like task simulation and collision detection but is time-consuming, demanding, and requires a complete digital model of the workspace~\cite{conventionalOfflineProgramming}. Another example is walk-through programming, where the robot's end effector is physically moved to the desired positions while the robot controller records these movements \cite{smallbusinessundermined}. 
    A concept based on this is programming by demonstration~\cite{reviewerArLFD}, in which the robot not only imitates the movements but also learns the tasks via machine learning to generalize these for similar situations~\cite{conventionalDemonstrationProgramming}.
    MR robot programming offers a solution to the limitations of these conventional methods since this technology provides an efficient hybrid solution that combines online and offline programming. It allows users to overlay the physical robot environment with a virtual robot replica, as depicted in Figure \ref{figureTeaser},  enabling them to program the virtual model first and simulate its movements. This process helps identify potential problems, such as collisions, early on. Once the immersive simulation runs smoothly, the physical robot can reliably execute the programmed tasks without having to create a detailed digital model of the environment.

    Research findings indicate that MR programming is more effective and efficient than conventional methods \cite{BaseRobotMR, BaseRobotMR2}. An analysis of recent publications in this field shows that the interaction is performed using gesture control in most cases \cite{BaseRobotMR3, exampleMRwithHoloGesture2, programmingWithHololensSlower, exampleMRwithHoloGesture3, BaseRobotMR3}. This involves the use of a remote manipulation interaction technique for setting waypoints \cite{BaseRobotMR, BaseRobotMR2, relatedworkVR2}. However, this method is less accurate than direct manipulation due to hand instability, inaccurate tracking systems, and difficulty mapping the user's movements to virtual objects \cite{reasonBadHandtracking1, reasonBadHandtracking2}. In addition, studies have shown that the performance of gesture-based interaction is not as good as controller-based interaction \cite{betterControllerInteractionThanHandtracking1, betterControllerInteractionThanHandtracking2} while using tangible objects leads to better results \cite{tangiblePositiveEffect1, tangiblePositiveEffect2, tangiblePositiveEffect3, EodMR2}.

    In this paper, we investigate the potential of controller-based MR robot programming by comparing the following three methods: 1) \CCJ, in which the user moves the virtual robot using the thumbsticks and buttons of a controller, 2) \CDC, in which the position and orientation of the controller (6-DoF) is transferred to the virtual end effector, and 3) \CDCwG, which extends the \CDC method with a mechanical gripper identical to that of the robot, enabling gripping and releasing of physical objects. These three methods are evaluated in a study ($n=30$) to determine their effectiveness and efficiency. A fast and uncomplicated robot programming solution for people with no previous experience or knowledge in human-robot interaction can accelerate the widespread introduction of robot technologies in small and medium-sized companies~\cite{smallbusinessundermined}.

\section{RELATED WORK}
    Yang \etal \cite{relatedworkVR1} compared a video see-through MR robot programming system with an online programming method. The MR method used an Oculus Rift HMD and a physical handheld controller for interaction. In the online method, participants could choose between manual leading and pendant teaching. Both methods were evaluated in a within-subjects study ($n=16$) in which participants completed two complex programming tasks. The results showed that participants were faster, set fewer waypoints, and had less cognitive load when using the MR method.
    Gadre \etal \cite{BaseRobotMR} implemented an MR interface for robot programming using a Microsoft HoloLens with gesture interaction and compared it to a conventional offline 2D programming method (monitor, keyboard, and mouse). Participants had to perform two pick-and-place tasks in a within-subjects study ($n=20$). The analysis showed that the MR method was faster, easier, more natural, and less work-intensive than the conventional method. 
    Quintero \etal \cite{BaseRobotMR2} investigated an MR robot programming method by comparing it with a traditional kinesthetic teaching method. In a study ($n=10$), the participants performed free space and contact surface trajectory tasks. The interaction was performed via hand tracking using a Microsoft HoloLens and MYO armband. Results of a study showed that the MR method took less time, performed better, and required less physical effort but more mental effort. 
    Pizzagalli \etal \cite{baseARVRcompared} compared the two MR methods, AR (Microsoft HoloLens) and VR (Oculus Quest), with the online programming method teach pendant. Both MR methods were operated via gesture interaction. The results of a within-subjects study ($n=21$) showed that both MR methods were faster, easier to use, and less demanding for users than the conventional programming method.
    Dengxiong \etal \cite{relatedworkVR4} proposed a self-supervised 6-DoF grasp pose detection system through an MR teleoperation framework that efficiently learns useful grasp strategies from human demonstrations without requiring explicit grasp pose annotations. Thus, this method does not require precise grasp pose monitoring, which offers advantages in constrained environments. In real-world experiments, the proposed system demonstrated the ability to successfully grasp unknown objects after only three demonstrations.

\section{METHODS}
    In this section, we describe our novel robot programming approaches.
    As an apparatus for three methods, we use the Meta Quest 3\footnote{https://www.meta.com/de/quest/quest-3/} MR head-mounted display (HMD) as an interface. An Annin Robotics AR4 open-source robotic arm\footnote{https://www.anninrobotics.com/}, which is a desktop-sized 6-DoF industrial robot. A TCP relay server, running on a notebook, serves as the communication interface between the two components. In addition, a self-designed and 3D-printed controller extension is described in more detail in Section \ref{gripperControlExplained}.
    
    All three programming methods have the same basic functions for simplifying usability. This includes rendering the virtual robot based on the specified end effector position and orientation by performing real-time inverse kinematics calculations. All methods use a waypoint-based programming approach, where users define key positions and orientations that form the robot's trajectory. If the movements of the end effector overstretch the physical joint limits, the corresponding joint is visually highlighted. The robot's working envelope is visualized via a transparent shape that looks like a sphere so the user can assess its range. Furthermore, the trajectory of the end effector is calculated and visualized using geometric path planning. It is also possible to simulate this trajectory during programming, which has the advantage that the user can safely see the movements at close range and does not have to maintain a safe distance as required for a physical robot. 
    \begin{figure}[thpb]
      \centering
      \includegraphics[width=\linewidth]{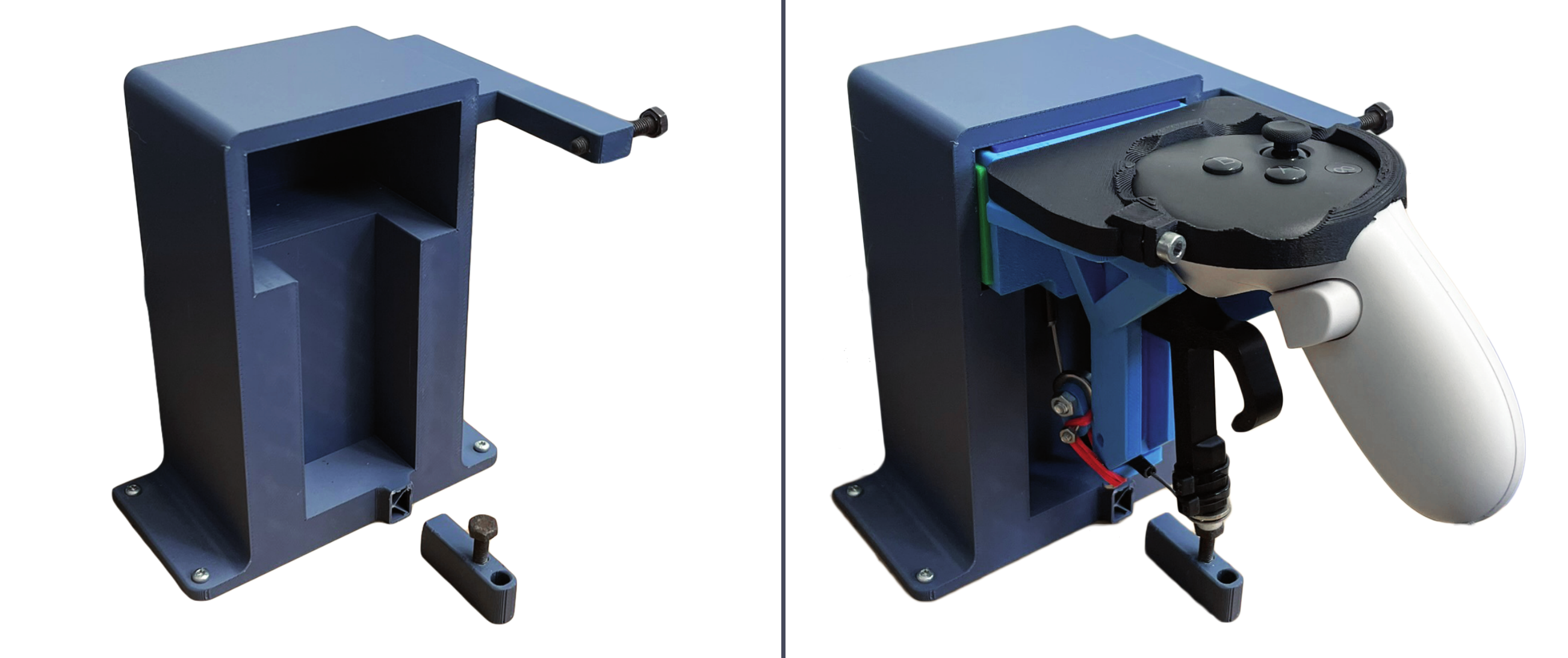}
      \caption{Construciton for synchronization of the virtual and physical robot bases. It is executed by inserting the 3D-printed gripper extension into the 3D-printed opening case.}
      \label{figureCalibration}
    \end{figure}
    The virtual and physical robot bases are synchronized, so the movements in the augmented simulation match those of the physical robot. This precise synchronization is executed by inserting the 3D-printed gripper extension into a 3D-printed opening (see Figure \ref{figureCalibration}) on the robot base. As the orientation and positioning of this opening relative to the physical robot and the tracked gripper controller are known, the virtual model is calibrated by executing a function with a controller button. 
    The software and its functions are implemented with Unity\footnote{https://unity.com/}.

\subsection{\NCCJ}
    The \CCJ condition is used as a baseline since it reflects the conventional teach pendant method~\cite{conventionalOnlineTeachPendants}. This allows the user to control the virtual end effector in two separately selectable modes. In Position Mode, the left joystick controls the end-effector's X- and Y-coordinates, while the right joystick controls its Z-coordinate. In Rotation Mode, the left joystick changes the pitch and yaw of the end-effector, while the right joystick modifies the roll angle. Modes can be switched using a controller button, while another button is used to program the defined waypoints. The gripper state can be changed using the grip trigger of a controller.

\subsection{\NCDC}
    The \CDC method allows the virtual end effector to follow the position and orientation of the Meta Quest controller in real-time. The waypoint can be captured with the controller button when the user reaches the desired position, orientation, and grasp status. As in the \CCJ condition, the state of the gripper can be changed by pressing the controller's grip trigger. %

\subsection{\NCDCwG} \label{gripperControlExplained}
\begin{figure}[thpb]
      \centering
      \includegraphics[width=\linewidth]{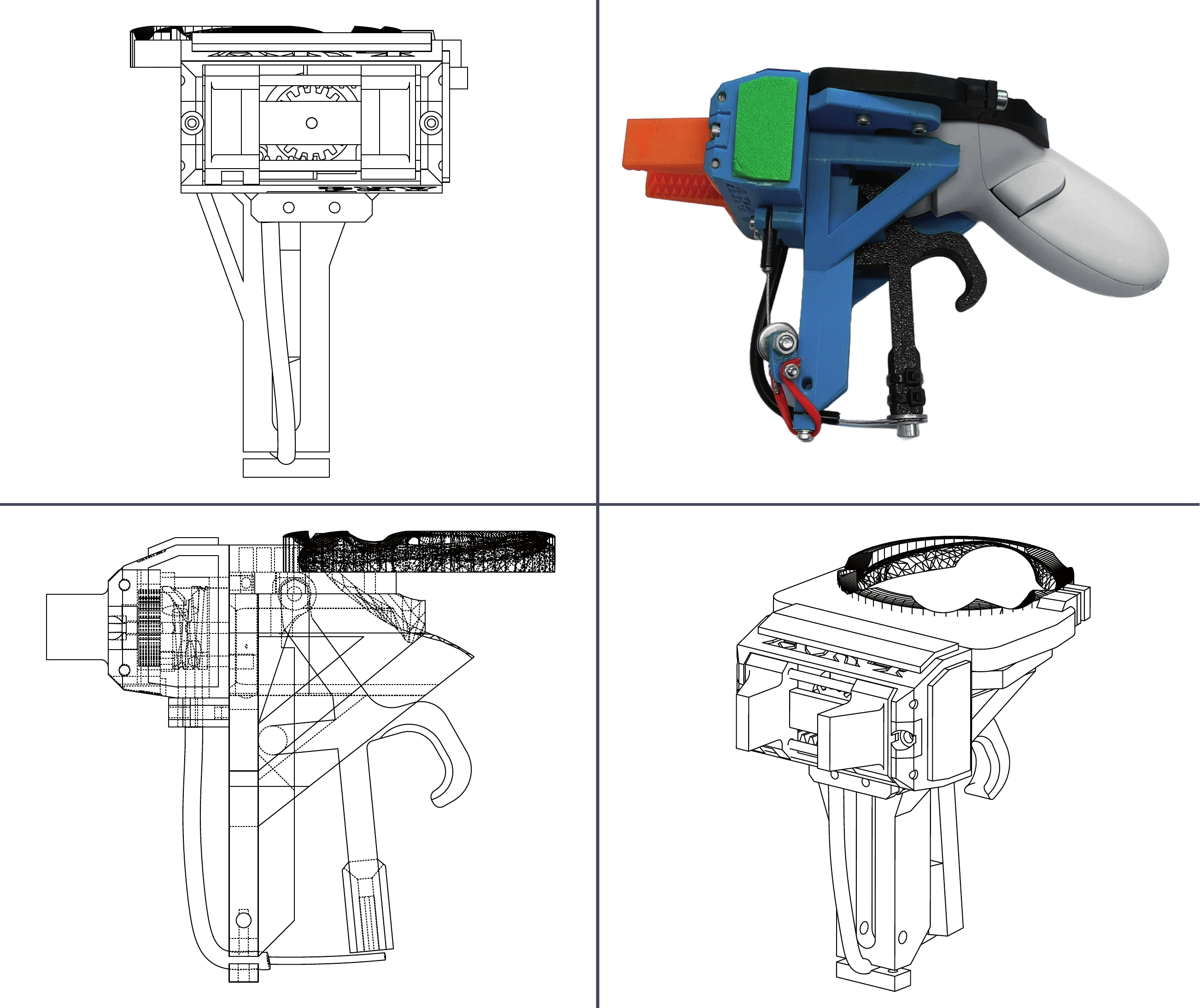}
      \caption{Visualization of the gripper used for condition \CDCwG.}
      \label{figureGripper}
    \end{figure}

    In the \CDCwG method, as in the \CDC method, the end effector's real-time position and orientation correspond with the controller's. The way objects are gripped and released differs here, as the Meta Quest controller is modified with a 3D-printed gripper extension shown in Figure \ref{figureGripper}. This extension is a tangible interface that transmits gripper states and provides haptic feedback when interacting with target objects. It allows the user to grip, move, and release physical objects while programming. This approach offers a conceptual advantage, although it has not been investigated in this contribution and is currently only a potential benefit, as the haptic force feedback may assist in handling fragile or deformable objects. All this is possible as the design of the 3D-printed gripper is identical to that of the servo gripper of the AR4 robot. The Gripper Controller is operable with a single hand, and its jaws are interchangeable for diverse pick-and-place tasks.

\section{EVALUATION}
\subsection{Study Design}
    We conducted a within-subjects study design to prevent variations caused by individual differences and to obtain a final evaluation of all programming methods. Using the three methods (\CCJ, \CDC, and \CDCwG) all participants completed the three levels of the independent variable. As dependent variables, we collected the participants' programming duration, subjective ratings for assessing the User Experience using the short version of the User Experience Questionnaire (UEQ-S) \cite{UeqShort}, the Workload using the Raw NASA Task Load Index Questionnaire (NASA-TLX) \cite{NasaTlx}, and specific questions for the feasibility, natural usability, precision, and overall assessment. 
    In all three methods, participants completed two tasks to assess their performance based on programming errors (\eg, collisions, incorrect grab/release of objects, or skipped programming steps). These errors were captured through software data logging and documented observations by the study supervisor.

\subsection{Procedure}
\begin{figure}[thpb]
      \centering
      \includegraphics[width=\linewidth]{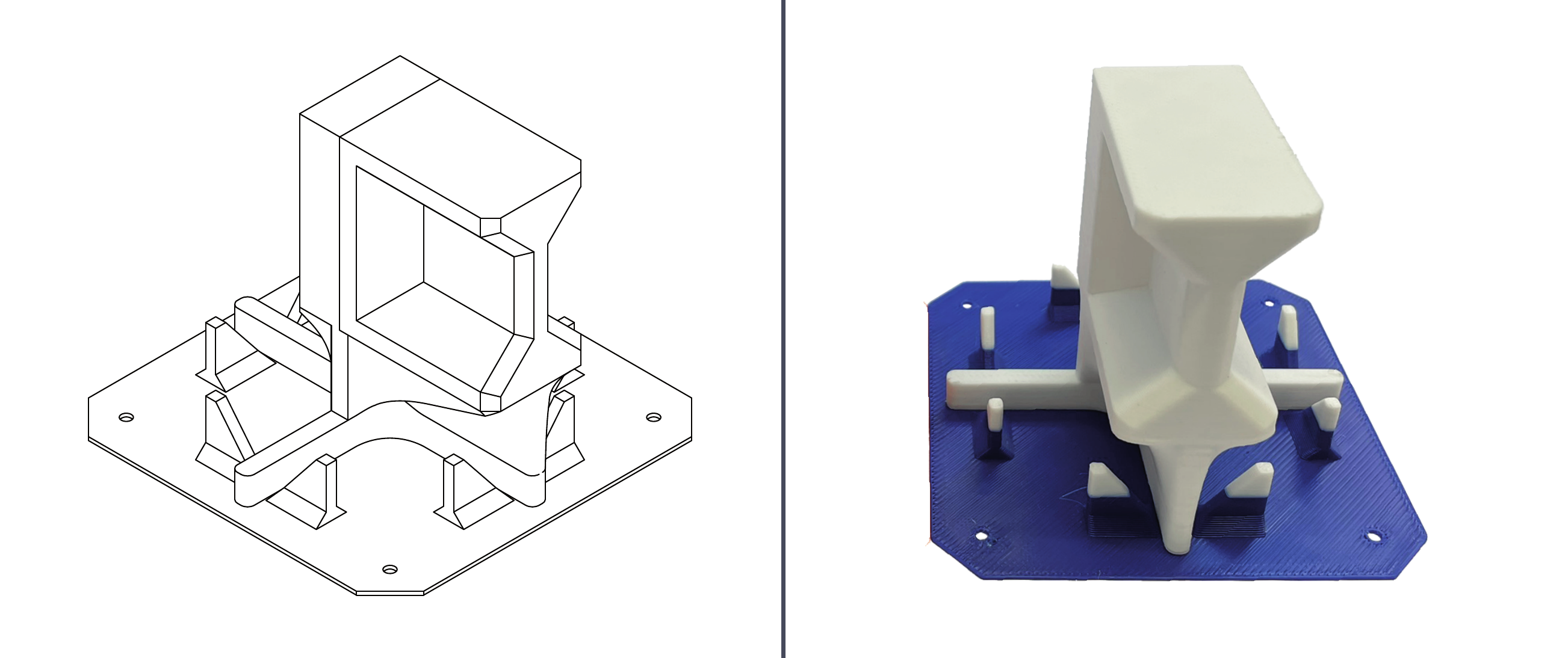}
      \caption{Target objects that were used for the two study tasks due to their self-correcting geometry.}
      \label{figureTargetObject}
    \end{figure}

    Initially, participants were briefed about the study and received a safety introduction (for the mixed reality and robotics hardware). Subsequently, they were asked to sign a consent form and provide demographic information. A video tutorial ($8:40$ min.) was presented to each participant to explain the functionality of the system and the three programming methods in an easy-to-understand and consistent manner. 
    Participants then completed the three programming methods following the identical procedure for each method. First, they had five minutes to familiarize themselves with the system and the programming method by moving the robot arm into various positions and orientations, operating the gripper, and programming the robot states. Afterward, there were two programming tasks with increasing levels of difficulty. We used self-developed \enquote{target objects}, which were designed to be easy to pick up and place down by employing self-correcting geometry like big chamfers (see Figure \ref{figureTargetObject}).

    \textbf{Task 1}: The first task (pick-and-place) was designed to be relatively simple, involving the action of grabbing a \enquote{target object} and dropping it into a cardboard box. This task was chosen to familiarize the participants with the programming method and to test their ability to carry out movements.
    
    \textbf{Task 2}: The second task involved swapping two \enquote{target objects}. There were three platforms, with one \enquote{target object} on both the left and right platforms. Participants could choose which \enquote{target object} to temporarily place on the third empty platform in the middle to complete the task. This procedure was more complex, as it required three grabbing and placing operations and, therefore, more programming steps. The collision risk was also higher.

    After completing the two tasks, participants completed a questionnaire that assessed their experience with the programming method. This procedure was repeated for each programming method. %
    Upon completing the third programming method, participants completed an additional questionnaire that focused on their overall experience of all conditions.
    To avoid \enquote{order effects} that might influence participants' performance, we counterbalanced the order of the three conditions according to the Latin Square. A live stream was used to monitor the participants' perspectives during the study, providing real-time feedback and assistance in case of problems. With a time limit of 5 minutes for each task, the study took approximately 60 minutes.

\subsection{Participants}
    The study was conducted with 30 participants (25 male, 5 female), aged between 19 and 38 years ($M = 25.17, SD = 4.29$). All were academics, most from a technical field.
    On a scale ranging from 1 =  \enquote{strongly disagree} to 5 =  \enquote{strongly agree}, they rated themselves as experienced in robot programming ($M = 2.63, SD = 1.73$) and MR ($M = 2.93, SD = 1.70$).

\section{Results}
    For parametric statistical data analysis ($\alpha$ = $0.05$), one-way repeated measures analysis of variance (ANOVA) and post-hoc paired t-tests with Bonferroni-Holm correction were used. Degrees of freedom were corrected in cases where Mauchly's Test indicated that the sphericity assumption was violated. Effect sizes were reported using $\eta_{p}^{2}$ (Partial Eta Squared): small ($> .01$), medium ($> .06$), and large ($> .14$).

\subsection{Task Completion Time}
    The statistical analysis of the duration (in seconds) required by the participants to complete the two tasks in the three programming methods is as follows.  
    
    \textbf{Duration task 1}: The Greenhouse-Geisser ($\epsilon = .54$) correction was used since Mauchly's Test indicated a violation of the sphericity ($\chi^{2}(2) = 53.612, p < .001$) assumption. This resulted in $F(1.08,31.307) = 54.477, p < .001, \eta_{p}^{2} = .653$. 
    The post-hoc analysis showed that participants completed the first task significantly faster in the conditions \CDCwG ($M=22.84, SD=10.33, p < .001$) and \CDC ($M=27.23, SD=17.64, p < .001$) compared to \CCJ ($M=122.53, SD=74.84$).
    
    \textbf{Duration task 2}: For the duration of the more complex task, a Greenhouse-Geisser ($\epsilon = .70$) correction was also applied due to a sphericity ($\chi^{2}(2) = 15.68, p < .001$) violation showing a significant effect $F(1.4,40.594) = 52.615, p < .001, \eta_{p}^{2} = .645$.
    The \CDCwG ($M=74.12, SD=28.49$) and \CDC ($M=109.50, SD=57.00$) conditions were significantly ($p < .001$) faster than \CCJ ($M=228.97, SD=102.52$). The Methods \CDCwG  and \CDC duration also differed significantly ($p = .029$).

\subsection{User Experience}
    \tikzstyle{EDR}=[draw=black!25,line width=1pt,preaction={clip, postaction={pattern=north west lines, pattern color=black!10}}]

\begin{figure}[tpb]
        \centering
          \begin{tikzpicture}
            \begin{axis}[
                ybar,
                width=\linewidth,
                height=4.9cm,
                ylabel={User Experience},
                ymin=0, ymax=3.2,
                ytick={0,0.5,..., 2.5},
                ymajorgrids = true,
                major grid style={draw=myLightGrey},
                enlarge y limits={value=.1,upper},
                tickwidth=0pt, %
                enlarge x limits=true,
                tickwidth=0pt,
                symbolic x coords={PQ, HQ, O},
                x tick label style={font=\small,text width=1.6cm,align=center},
                xtick align=inside,
                xtick={PQ, HQ, O},
                enlarge x limits= 0.3,
                bar width=0.50cm,
            ]
            
                \addplot[fill=myRed, draw=none, opacity = 1.0, line width=0.0mm, error bars/.cd, 
                y dir=both, y explicit] 
                     coordinates { 
                        (PQ, 0.108) +- (0, 0.182)
                        (HQ, 0.892) +- (0, 0.202)
                        (O, 0.50) +- (0, 0.167)
                     };
                     \draw [{Bar[left]}-{Bar[right]}] (-23,338) -- (23,338)   node [midway,fill=white, inner sep = 0.2mm] {\tiny$\ast\kern-0.5em \ast\kern-0.5em\ast$};
                    \draw [{Bar[left]}-{Bar[right]}] (-23,308) -- (0.3,308)   node [midway,fill=white, inner sep = 0.2mm] {\tiny$\ast\kern-0.5em \ast\kern-0.5em\ast$};
                    \draw [{Bar[left]}-{Bar[right]}] (0.3,278) -- (023,278)   node [midway,fill=white, inner sep = 0.2mm] {\tiny$\ast\kern-0.5em \ast\kern-0.5em\ast$};
                    
                \addplot[fill=myGreen,  draw=none, opacity = 1.0, line width=0.0mm, error bars/.cd, 
                y dir=both, y explicit] 
                    coordinates { 
                        (PQ, 1.308) +- (0, 0.195)
                        (HQ, 1.90) +- (0, 0.171)
                        (O, 1.604) +- (0, 0.139)
                    };
                    \draw [{Bar[left]}-{Bar[right]}] (77,338) -- (123,338)   node [midway,fill=white, inner sep = 0.2mm] {\tiny$\ast\kern-0.5em \ast\kern-0.5em\ast$};
                    \draw [{Bar[left]}-{Bar[right]}] (77,308) -- (100.3,308)   node [midway,fill=white, inner sep = 0.2mm] {\tiny$\ast\kern-0.5em \ast\kern-0.5em\ast$};

                \addplot[fill=myBlue,  draw=none, opacity = 1.0, line width=0.0mm, error bars/.cd, 
                y dir=both, y explicit] 
                    coordinates { 
                        (PQ, 2.267) +- (0, 0.139)
                        (HQ, 2.20) +- (0, 0.139)
                        (O, 2.233) +- (0, 0.105)
                    };
                    \draw [{Bar[left]}-{Bar[right]}] (177,338) -- (223,338)   node [midway,fill=white, inner sep = 0.2mm] {\tiny$\ast\kern-0.5em \ast\kern-0.5em\ast$};
                    \draw [{Bar[left]}-{Bar[right]}] (177,308) -- (200.3,308)   node [midway,fill=white, inner sep = 0.2mm] {\tiny$\ast\kern-0.5em \ast\kern-0.5em\ast$};
                    \draw [{Bar[left]}-{Bar[right]}] (200.3,278) -- (223,278)   node [midway,fill=white, inner sep = 0.2mm] {\tiny$\ast\kern-0.5em \ast\kern-0.5em\ast$};
                    
            \end{axis}
            \matrix [below] at (current bounding box.south) {
              \filldraw [myRed, opacity = 1.0, scale=0.4] plot coordinates {(0,0)(0.8,0)(0.8,0.4)(0,0.4)} -- cycle; & \node[anchor=base]{\small \NCCJ}; &
              \filldraw [myGreen, opacity = 1.0, scale=0.4] plot coordinates {(0,0)(0.8,0)(0.8,0.4)(0,0.4)} -- cycle; & \node[anchor=base]{\small \NCDC}; &
              \filldraw [myBlue, opacity = 1.0, scale=0.4] plot coordinates {(0,0)(0.8,0)(0.8,0.4)(0,0.4)} -- cycle; & \node[anchor=base]{\small \NCDCwG}; \\
            };
        \end{tikzpicture}
      \caption{The means of the UEQ ratings for each condition. Scales are (PQ) Pragmatic Quality, (HQ) Hedonic Quality, and (O) Overall, ranging from -3 to +3. Error bars represent the standard error.}
      \label{fig:ueqResults}
    \end{figure}
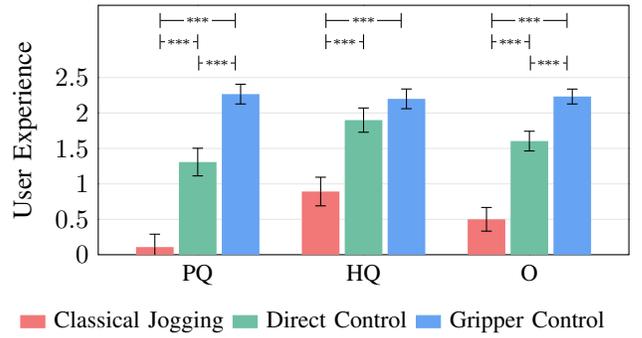

    Participants rated the eight items of the UEQ-S questionnaire using a 7-point Likert scale, ranging from -3 = \enquote{horribly bad} to 3 = \enquote{extremely good}. Four of these items represent the pragmatic quality, and the other four are the hedonic quality (see Figure \ref{fig:ueqResults}). 

    \textbf{Pragmatic quality} showed a significant effect $F(2,58) = 36.971, p < .001, \eta_{p}^{2} = .560$. It involves factors such as perspicuity, efficiency, and dependability, so the interaction quality was significantly ($p < .001$) higher in the \CDCwG ($M=2.27, SD=0.76$) and \CDC ($M=1.31, SD=1.07$) conditions compared to the \CCJ ($M=0.11, SD=1.0$) conditions. The \CDCwG was also significantly ($p < .001$) better than the \CDC condition. 
    
    \textbf{Hedonic quality} also showed a significant main effect $F(2,58) = 34.023, p < .001, \eta_{p}^{2} = .540$. This quality includes stimulation and novelty, so the pleasure or fun of the product was significantly ($p < .001$) better rated for the \CDCwG ($M=2.20, SD=0.76$) and \CDC ($M=1.90, SD=0.93$) conditions compared to the \CCJ ($M=0.89, SD=1.11$). 
    
    \textbf{Overall} mean scores showed a clear difference in the statistical analysis $F(2,58) = 48.857, p < .001, \eta_{p}^{2} = .628$.
    The \CDCwG ($M=2.23, SD=0.58$) condition received significantly ($p < .001$) better ratings than the \CDC ($M=1.60, SD=0.76$), and \CCJ ($M=0.50, SD=0.91$). \CDC was also significantly ($p < .001$) better than \CCJ.

\subsection{Workload}
    \begin{figure*}[thpb]
        \centering
          \begin{tikzpicture}
            \begin{axis}[
                ybar,
                width=\linewidth,
                height=4.9cm,
                ymin=0.0, ymax=78,
                ylabel={Workload},
                ytick={0,20,..., 60},
                ymajorgrids = true,
                major grid style={draw=myLightGrey},
                enlarge y limits={value=.1,upper},
                tickwidth=0pt, %
                enlarge x limits=true,
                tickwidth=0pt,
                tick align=inside,
                symbolic x coords={MD, PD, TD, P, E, F, O},
                xtick align=inside,
                xtick={MD, PD, TD, P, E, F, O},
                bar width=13pt,%
                no markers,
            ]   
                \addplot[fill=myRed,  draw=none, opacity = 1.0, line width=0.0mm, error bars/.cd, y dir=both, y explicit] 
                     coordinates { 
                        (MD, 57.833) +- (0, 4.386)
                        (PD, 22.00) +-  (0, 4.122)
                        (TD, 31.167) +- (0, 4.591)
                        (P, 33.167) +-  (0, 4.725)
                        (E, 55.167) +-  (0, 4.278)
                        (F, 34.333) +-  (0, 4.470)
                        (O, 38.944) +-  (0, 2.839)
                     };

                \addplot[fill=myGreen,  draw=none, opacity = 1.0, line width=0.0mm, error bars/.cd, y dir=both, y explicit] 
                    coordinates { 
                        (MD, 38.667) +- (0, 4.301)
                        (PD, 19.00) +- (0, 3.831)
                        (TD, 30.833) +- (0, 3.761)
                        (P, 38.333) +- (0, 5.308)
                        (E, 31.167) +- (0, 4.239)
                        (F, 19.333) +- (0, 2.772)
                        (O, 29.556) +- (0, 2.488)
                    };
                \addplot[fill=myBlue,  draw=none, opacity = 1.0, line width=0.0mm, error bars/.cd, y dir=both, y explicit] 
                coordinates { 
                        (MD, 26.50) +- (0, 3.738)
                        (PD, 28.0) +- (0, 3.818)
                        (TD, 27.833) +- (0, 4.205)
                        (P, 28.833) +- (0, 5.747)
                        (E, 25.833) +- (0, 4.041)
                        (F, 17.167) +- (0, 4.386)
                        (O, 25.694) +- (0, 2.851)
                };
            \draw [{Bar[left]}-{Bar[right]}]  (-28.40,810) -- (28.40,810)  node [midway, fill=white, inner sep = 0.2mm] {\tiny$\ast\kern-0.5em \ast\kern-0.5em\ast$}; 
             
            \draw [{Bar[left]}-{Bar[right]}]  (-28.40,735) -- (0.40,735)  node [midway, fill=white, inner sep = 0.2mm] {\tiny$\ast\kern-0.5em \ast\kern-0.5em\ast$};  

            \draw [{Bar[left]}-{Bar[right]}]  (-0.34,660) -- (28.40,660)  node [midway, fill=white, inner sep = 0.2mm] {\tiny$\ast\kern-0.18em \ast$};

            \draw [{Bar[left]}-{Bar[right]}]  (371.60,810) -- (428.95,810)  node [midway, fill=white, inner sep = 0.2mm] {\tiny$\ast\kern-0.5em \ast\kern-0.5em\ast$};  
            
            \draw [{Bar[left]}-{Bar[right]}]  (371.60,735) -- (400.55,735)  node [midway, fill=white, inner sep = 0.2mm] {\tiny$\ast\kern-0.5em \ast\kern-0.5em\ast$};  
            
            \draw [{Bar[left]}-{Bar[right]}]  (471.55,810) -- (528.70,810)  node [midway, fill=white, inner sep = 0.2mm] {\tiny$\ast\kern-0.18em \ast$};  

            \draw [{Bar[left]}-{Bar[right]}]  (471.55,735) -- (500.50,735)  node [midway, fill=white, inner sep = 0.2mm] {\tiny$\ast\kern-0.18em \ast$};

            \draw [{Bar[left]}-{Bar[right]}]  (571.35,810) -- (628.50,810)  node [midway, fill=white, inner sep = 0.2mm] {\tiny$\ast\kern-0.5em \ast\kern-0.5em\ast$};  
            
            \draw [{Bar[left]}-{Bar[right]}]  (571.35,735) -- (600.30,735)  node [midway, fill=white, inner sep = 0.2mm] {\tiny$\ast\kern-0.5em \ast\kern-0.5em\ast$};

            \end{axis}
            \matrix [below] at (current bounding box.south) {
              \filldraw [myRed, opacity = 1.0, scale=0.4] plot coordinates {(0,0)(0.8,0)(0.8,0.4)(0,0.4)} -- cycle; & \node[anchor=base]{\small \NCCJ}; &
              \filldraw [myGreen, opacity = 1.0, scale=0.4] plot coordinates {(0,0)(0.8,0)(0.8,0.4)(0,0.4)} -- cycle; & \node[anchor=base]{\small \NCDC}; &
              \filldraw [myBlue, opacity = 1.0, scale=0.4] plot coordinates {(0,0)(0.8,0)(0.8,0.4)(0,0.4)} -- cycle; & \node[anchor=base]{\small \NCDCwG}; \\
            };
        \end{tikzpicture}
      \caption{Mean scores of the NASA-RTLX ratings ranging from 0 [very low] to 100 [very high]. The scales are (MD) Mental demand, (PD) Physical demand, (TD) Temporal demand, (P) Performance, (E) Effort, (F) Frustration, and (O) Overall. Error bars indicate the standard error.}
      \label{fig:nasaResults}
    \end{figure*}
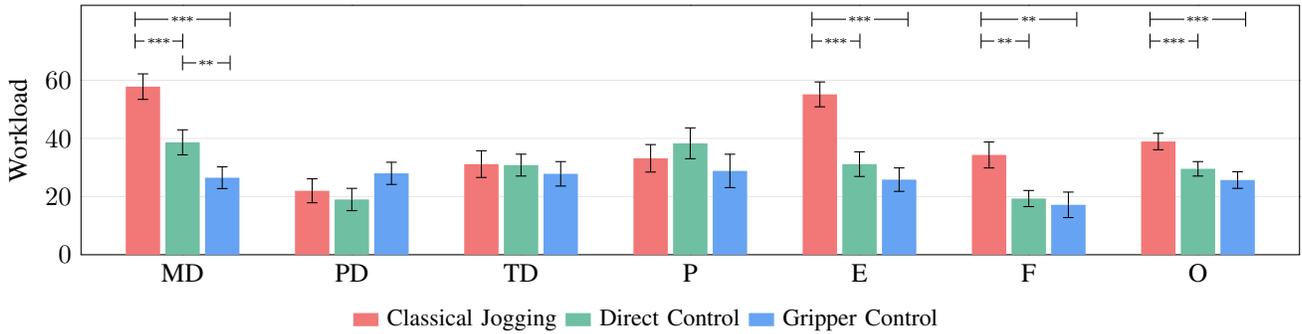

    Perceived workload ratings (where lower ratings indicate a positive assessment and higher ratings a negative one) are presented in Figure \ref{fig:nasaResults} and statistically analyzed as follows. Excluding the NASA-TLX items, \textbf{Physical demand}, \textbf{Temporal demand}, and \textbf{Performance}, the repeated measures ANOVA showed an effect on the remaining items. 
    
    \textbf{Mental demand} $F(2,58) = 37.826, p < .001, \eta_{p}^{2} = .566$ was significantly the lowest for condition \CDCwG ($M=26.50, SD=20.48$) compared to \CDC ($M=38.67, SD=23.56, p = .001$) and \CCJ ($M=57.83, SD=24.02, p < .001$). %
    An effect ($p < .001$) was also between \CDC and \CCJ.
    
     \textbf{Effort} Mauchly's Test indicated that the assumption of sphericity had been violated ($\chi^{2}(2) = 7.796, p = .02$), so degrees of freedom are corrected using Huyn-Feldt estimation of sphericity ($\epsilon = .845$). Results showed a significant effect $F(1.689,48.987) = 29.844, p < .001, \eta_{p}^{2} = .507$. Compared to method \CCJ ($M=55.17, SD=23.43$), participants had to work significantly less with methods \CDCwG ($M=25.83, SD=22.13$) and \CDC ($M=31.17, SD=23.22$) to accomplish their level of performance. 
    
     \textbf{Frustration} $F(2,58) = 7.416, p = .001, \eta_{p}^{2} = .204$ was significantly lower with the methods \CDCwG ($M=17.17, SD=24.02, p = .002$) and \CDC ($M=19.33, SD=15.19, p = .006$) compared to \CCJ ($M=34.33, SD=24.49$).
    
    \textbf{Overall} mean scores of all six NASA-TLX items showed a clear difference in the statistical analysis $F(2,58) = 17.947, p < .001, \eta_{p}^{2} = .382$, as the workload for the \CDCwG ($M=25.69, SD=15.62$) and \CDC ($M=29.56, SD=13.63$) methods was significantly ($p < .001$) lower than \CCJ ($M=38.94, SD=15.55$).

\subsection{Specific Questions}
    The statistical analysis of the specific questions, which were rated from 1 = \enquote{strongly disagree} to 7 = \enquote{strongly agree}, showed significant effects.
    
    \textbf{Q1}: \enquote{I was able to fulfill the tasks without any problems.} $F(2,58) = 10.140, p < .001, \eta_{p}^{2} = .259$. 
    The \CDCwG ($M=6.27, SD=0.83$) condition significantly achived the best results compared to \CDC ($M=5.07, SD=1.44, p < .001$) and \CCJ ($M=5.13, SD=1.14, p < .001$).

    \textbf{Q2}: \enquote{The programming technique felt natural to use.} $F(2,58) = 28.498, p < .001, \eta_{p}^{2} = .496$. Conditions \CDCwG ($M=6.20, SD=1.06, p < .001$) and \CDC ($M=5.57, SD=1.22, p < .001$) were rated significantly better than \CCJ ($M=3.73, SD=1.51$).

    After completing all three programming methods, the participants evaluated them using the following questions.
    
    \textbf{Q3}: \enquote{I was able to program the robot precisely.} since Mauchly's Test indicated a violated assumption of sphericity ($\chi^{2}(2) = 13.033, p = .001$), the degrees of freedom are corrected using Huyn-Feldt ($\epsilon = .757$) resulting in $F(1.515,43.935) = 11.574, p < .001, \eta_{p}^{2} = .285$.
    The post-hoc test indicated that \CDCwG ($M=6.37, SD=0.77$) was significantly better than \CDC ($M=5.03, SD=1.22, p < .001$) and \CCJ ($M=5.13, SD=1.43, p < .001$).   

    \textbf{Q4}: \enquote{Overall, I would rate the method.}, $F(2,58) = 21.253, p < .001, \eta_{p}^{2} = .423$. The ratings of the condition \CDCwG ($M=6.27, SD=0.98$) were significantly higher compared to \CDC ($M=5.40, SD=1.13, p = .010$) and \CCJ ($M=4.17, SD=1.58, p < .001$). Also, the conditions \CDC and \CCJ revealed an effect ($p < .001$).

\subsection{Task Performance}
    \usepgfplotslibrary{groupplots}
\usepgfplotslibrary{statistics}
    \pgfplotsset{
        box style/.style={
            #1,
            solid,
            fill=#1,
            fill opacity=1.0,
            mark=o,
            mark size=1.00pt,
            mark options={myGrey},
            boxplot={
                draw position={1/6 + floor(\plotnumofactualtype/5) + 1/6*mod(\plotnumofactualtype,5)},
                box extend=0.15,
                every whisker/.style={black},
                every median/.style={black},
            },
        },
        mark style/.style={
            #1,
            mark=*,
            only marks,
            table/x expr={1/6 + floor(\plotnumofactualtype/5) + 1/6*mod(\plotnumofactualtype,5)},
        },
    }

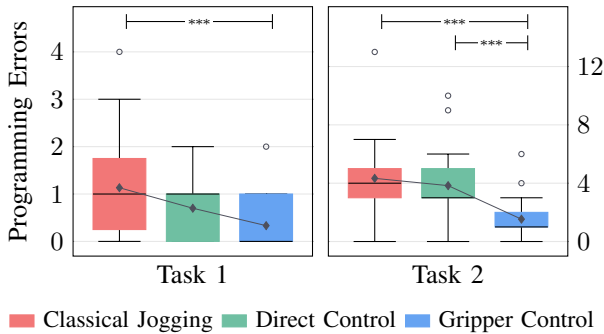
\begin{figure}[thpb]
    \centering
	\begin{tikzpicture}
		\begin{groupplot}[group 
		style={group size=2 by 1,horizontal sep=6pt},
		width=\linewidth/1.8125, %
        height=4.9cm,
        ymajorgrids = true,
        major grid style={draw=myLightGrey},
        enlarge y limits={value=.1,upper},
        enlarge x limits=true,
        tickwidth=0pt,
        tick align=inside,
        xtick={0,...,2},
        x tick label as interval,
        x tick label style={text width=3.6cm,align=center},
        boxplot/draw direction=y,
		ymajorgrids = true,major grid style={draw=myLightGrey},
        ]

		\nextgroupplot[ylabel={Programming Errors}, xlabel={Task 1}, ymin=-0.31, ymax=4.475,
        ytick={0, ..., 4},] 
        
                \addplot+ [
                    box style=myRed,
                    boxplot prepared={
                        lower whisker=0.00,
                        lower quartile=0.25,
                        median=1.0,
                        upper quartile=1.75,
                        upper whisker=3.00,
                    },boxplot/box extend= 0.12] coordinates {(0,4)};
                  
                \addplot+ [
                    box style=myGreen,
                    boxplot prepared={
                        lower whisker=0.00,
                        lower quartile=0.00,
                        median=1.00,
                        upper quartile=1.00,
                        upper whisker=2.00,
                    },boxplot/box extend= 0.12] coordinates {};
                    
                \addplot+ [
                    box style=myBlue,
                    boxplot prepared={
                        lower whisker=0.00,
                        lower quartile=0.00,
                        median=0.00,
                        upper quartile=1.0,
                        upper whisker=1,
                    },boxplot/box extend= 0.12] coordinates {(0,2)};

                \addplot [color = myGrey,mark=diamond*, solid, mark size=1.5pt, line width=0.100mm,] coordinates {
                    (0.1666,1.133) 
                    (0.3334,0.70)
                    (0.5,0.333)
                    };

                \draw [{Bar[left]}-{Bar[right]}]  (74.6,495) -- (408.8,495)  node [midway,fill=white, inner sep = 0.2mm] {\tiny$\ast\kern-0.5em \ast\kern-0.5em\ast$};

		\nextgroupplot[yticklabel pos=right, xlabel={Task 2}, ymin=-1.0, ymax=14.55,
        ytick={0,4,..., 12},]

                \addplot+ [
                    box style=myRed,
                    boxplot prepared={
                        lower whisker=0.00,
                        lower quartile=3.00,
                        median=4.0,
                        upper quartile=5.0,
                        upper whisker=7.00,
                    },boxplot/box extend= 0.12] coordinates {(0,13)};
                  
                \addplot+ [
                    box style=myGreen,
                    boxplot prepared={
                        lower whisker=0.00,
                        lower quartile=3.00,
                        median=3.00,
                        upper quartile=5.00,
                        upper whisker=6.00,
                    },boxplot/box extend= 0.12] coordinates {(0,10)(0,9)};
                    
                \addplot+ [
                    box style=myBlue,
                    boxplot prepared={
                        lower whisker=0.00,
                        lower quartile=1.00,
                        median=1.00,
                        upper quartile=2.00,
                        upper whisker=3.00,
                    },boxplot/box extend= 0.12] coordinates {(0,6)(0,4)};
                
                \addplot [color = myGrey,mark=diamond*, solid, mark size=1.5pt, line width=0.100mm,] coordinates {
                    (0.1666,4.333) 
                    (0.3334,3.833)
                    (0.5,1.533)
                    };

                \draw [{Bar[left]}-{Bar[right]}]  (74.6,160.8) -- (408.8,160.8)  node [midway,fill=white, inner sep = 0.2mm] {\tiny$\ast\kern-0.5em \ast\kern-0.5em\ast$};
                \draw [{Bar[left]}-{Bar[right]}]  (242.1,146.3) -- (408.8,146.3)  node [midway,fill=white, inner sep = 0.2mm] {\tiny$\ast\kern-0.5em \ast\kern-0.5em\ast$};

		\end{groupplot}
        \matrix [below] at (current bounding box.south) {
              \filldraw [myRed, opacity = 1.0, scale=0.4] plot coordinates {(0,0)(0.8,0)(0.8,0.4)(0,0.4)} -- cycle; & \node[anchor=base]{\small \NCCJ}; &
              \filldraw [myGreen, opacity = 1.0, scale=0.4] plot coordinates {(0,0)(0.8,0)(0.8,0.4)(0,0.4)} -- cycle; & \node[anchor=base]{\small \NCDC}; &
              \filldraw [myBlue, opacity = 1.0, scale=0.4] plot coordinates {(0,0)(0.8,0)(0.8,0.4)(0,0.4)} -- cycle; & \node[anchor=base]{\small \NCDCwG}; \\
            };

	\end{tikzpicture}
    \caption{Errors that occurred during the two robot programming tasks. The connected diamonds represent the mean values, and the dots are the outliers.}
    \label{fig:performance}
\end{figure}

    \textbf{Performance task 1}:
        The problem-free completion of task 1 was most frequently achieved with condition \CDCwG ($70.0\%$), followed by \CDC ($46.7\%$) and method \CCJ ($26.7\%$).
        As illustrated in Figure \ref{fig:performance}, there is a significant difference in the number of errors made by the participants $F(2,58) = 6.911, p = .002, \eta_{p}^{2} = .192$. 
        \CDCwG ($M=0.33, SD=0.55$) had significantly less errors compared to \CCJ ($M=1.13, SD=1.01$). \CDC ($M=0.70, SD=0.75$) revealed no statistical difference.
        
    \textbf{Performance task 2}:
        With condition \CDCwG ($23.3\%$), task 2 was performed most often without any problems, followed by \CDC ($6.7\%$) and \CCJ ($3.3\%$).
        The number of programming errors showed a statistical effect $F(2,58) = 30.910, p < .001, \eta_{p}^{2} = .516$. Compared to \CDC ($M=3.83, SD=2.29, p < .001$) and \CCJ ($M=4.33, SD=2.23, p < .001$), \CDCwG ($M=1.53, SD=1.57$) showed a significantly lower number of errors.

\section{DISCUSSION}
    The user study results are consistent with those of earlier works, in which the conventional method performed worst \cite{BaseRobotMR, relatedworkVR1, BaseRobotMR2}. It should be noted that the \CCJ method, which serves as a baseline, has certain advantages (\eg, augmented representation, path planning, trajectory simulation) compared to traditional methods such as the teach pendant. In spite of this, setting the position and orientation using separated axes proved to take much longer than 6-DOF interactions, as multiple mode changes and perspective changes were needed to define a robot pose.

    In contrast, the \CDC method performs significantly better due to its controller-based 6-DOF interaction, as shown by the results of the UEQ and NASA RTLX questionnaires and the recorded programming errors. Since this method differs from the \CCJ method only in its interaction technique, the results can only be attributed to the fact that moving the controller to the desired position requires less effort than moving it with the joysticks, which is supported by more natural usability (Q2).
    Compared to the existing hand-tracking-based MR approaches, the controller-based approach is more suitable due to its haptics and better tracking accuracy \cite{controllerBetterAccuracy}. 

    Overall, the results strongly support the use of the \CDCwG method. It is the most efficient since its completion time of the more complex and, therefore, more meaningful task 2 was significantly the fastest. The overall results of the UEQ questionnaire are also better and the mental effort is significantly the lowest. 
    This method is also the most efficient, as evidenced by the lowest number of programming errors in the relevant task 2 and the results of the specific questions Q1 and Q3. 

    Our study has some limitations. As mentioned by Yang \etal \cite{relatedworkVR1} regarding rendering quality, the imperfect occlusion between augmented and physical objects can lead to errors in depth perception, \eg, rendering virtual objects completely when parts of them are already inside or behind a physical object. In addition, using the gripper and controller buttons simultaneously can be challenging for people with small hands. The manual calibration is imperfect and can lead to misalignment between the digital twin and the real robot, which occurs when the user briefly removes the HMD from the head. This could be improved by a computer vision-based calibration, \eg, object or marker recognition. However, it is currently not possible to access the real-time pass-through video of the Quest 3. 
    
    A promising direction for future work would involve improving the \textit{Gripper Controller}, which holds potential for enhancements in both the construction (\eg, increased stiffness, reduced mechanical friction) and ergonomics (\eg, a longer trigger for more gripping force, or the use of a push interaction instead of a trigger). Additionally, the controller extension can be designed for the attachment of different robot arm extensions.

\section{CONCLUSIONS}
    This article presents three controller-based methods for MR robot programming. These are 1) \CCJ, where the virtually augmented end effector is moved using the thumbsticks of two controllers, 2) \CDC, in which the position of the end effector directly matches that of the controller, and 3) \CDCwG, which builds upon \CDC by extending the controller with an interactive 3D-printed gripper, allowing intuitive programming by directly gripping, moving, and placing physical objects. A within-subjects study ($n=30$) was conducted to compare these methods. Participants completed two tasks for each method. 
    The results demonstrate that the \CDCwG condition is the fastest, has the best user experience, the lowest mental demand, and the best subjective and objective task performance. Therefore, the \CDCwG method has the potential to significantly improve future robot programming.

\balance

\bibliographystyle{plain}
\bibliography{main}

\end{document}